\documentclass{webofc}

\usepackage[varg]{txfonts}   
\usepackage{hyperref}
\usepackage{url}
\hypersetup{colorlinks=true,citecolor=blue,urlcolor=blue,linkcolor=blue}
\begin{document}
\title{Simulating Charm Quarks in IP-Glasma Initial Stage and Quark-Gluon Plasma: A Hybrid Approach for charm quark phenomenology}
%
%

\author{
 \firstname{Charles} \lastname{Gale}\inst{1} \and
  \firstname{Sangyong} \lastname{ Jeon}\inst{1} \and
\firstname{Manu} \lastname{Kurian}\inst{2}\fnsep\thanks{\email{manukurian10406@gmail.com}} \and
  \firstname{Bj\"orn} \lastname{Schenke}\inst{3} \and
   \firstname{Mayank} \lastname{Singh}\inst{4} 
}

\institute{Department of Physics, McGill University, 3600 University Street, Montreal, QC, H3A 2T8, Canada    
\and
RIKEN BNL Research Center, Brookhaven National Laboratory, Upton, New York 11973, USA
\and
 Physics Department, Brookhaven National Laboratory, Upton, New York 11973, USA
\and
  Department of Physics and Astronomy, Vanderbilt University, Nashville, TN 37240, USA  
          }

\abstract{We present phenomenological findings on charm quark transport while including its energy loss in both pre-equilibrium and hydrodynamic stages of the evolution.
We employed the MARTINI event generator for the production and evolution of heavy quarks in the relativistic heavy-ion collisions. The sensitivity of the heavy meson nuclear modification factor and flow coefficient to the early stage of heavy-ion collisions and bulk medium evolution is analyzed for Pb+Pb collisions at 5.02 TeV. Our study provides insights into the interaction strength of charm quarks during the early phase and within the quark-gluon plasma.}
\maketitle
\section{Introduction}
\label{intro}

The accurate modeling of quark-gluon plasma evolution generated in relativistic nucleus-nucleus collisions relies on the understanding of the dynamics in the collision's early phase. Being generated during the early phase of the collisions, heavy quarks serve as effective probes of the early stages of the collision event. Heavy quark transport in the medium can therefore provide information about temperature, viscosity, and other transport properties. Efforts have been made to understand the Brownian transport of heavy-flavor particles in the quark-gluon plasma (QGP) and to estimate key experimental observables such as the nuclear suppression factor $R_{AA}$ and elliptic flow $v_2$ at RHIC and LHC energies. Despite the existence of many studies on heavy-flavor dynamics in an expanding viscous QGP medium and the estimation of non-equilibrium effects on the heavy quark transport coefficients, major challenges persist in the phenomenological study of heavy quarks. The effects of the initial stage and the pre-equilibrium evolution of heavy quarks in the initial background are usually ignored. 

We model relativistic heavy-ion collisions at LHC energies with a hybrid dynamical approach consisting of a fluctuating IP-Glasma initial state~\cite{Schenke:2012wb}, followed by the viscous hydrodynamic approach MUSIC~\cite{Schenke:2010nt}. We evolve charm quarks in both the IP-Glasma initial stage and the hydrodynamically expanding QGP phase using the MARTINI event generator~\cite{Young:2011ug}. By integrating charm quark energy loss in the fluctuating initial state, in addition to that in the QGP phase, we present the phenomenological implications of charm quark evolution in the early stages of heavy-ion collisions.
\section{Charm quark multi-stage evolution}
\label{sec-1}

\begin{figure}
\centering
\includegraphics[width=14cm,clip]{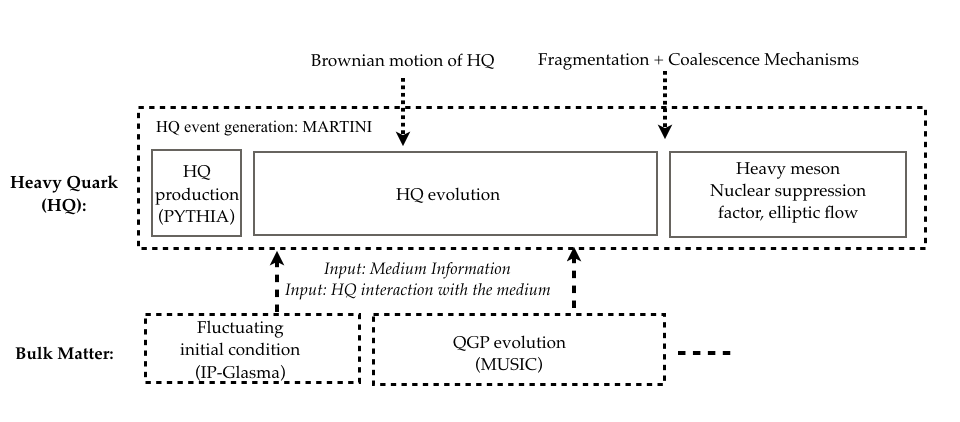}
\caption{Multi-stage model for heavy quark evolution in heavy-ion collisions}
\label{fig-1}      
\end{figure}
We describe heavy-quark evolution with a multi-stage model as described in figure~\ref{fig-1}. We initially generate charm quarks using PYTHIA8.2, which involves sampling the 6-dimensional momentum distributions of $Q\bar{Q}$ systems. The initial spatial distribution of charm quarks is obtained from the IP-Glasma initial state, with charm quarks randomly sampled on the background IP-Glasma. The precise determination of charm quark interactions in the expanding space-time geometry, which captures the gluon fields in the fluctuating initial state, is non-trivial and not yet fully developed. The phenomenological implication of the impact of charm quark evolution in the pre-equilibrium stage is an interesting aspect to explore.
In the present analysis, as a first step toward capturing the impact of the fluctuating initial state on charm quark evolution, we approximate the initial state as an evolving QCD medium dominated by gluons. We assume that the gluons form a thermalized medium with a conformal equation of state where a temperature can be defined. The transport of heavy quarks is studied as Brownian motion in such a medium, where the interaction strength of the quarks in the medium can be measured in terms of the appropriate diffusion coefficient.

The next phase is the evolution of the charm quark in the hydrodynamically expanding QGP. Relativistic second-order viscous hydrodynamics is employed to describe the evolution of the QGP medium. The change in charm quark momentum, $dp_i$, over a time interval $dt$, can be quantified numerically using the Langevin equations in the local rest frame of the medium as follows~\cite{Moore:2004tg},
\begin{align}
   &dp_i=-\eta(p_i) p_i\, dt+ \xi_i ({\bf p})\, {dt},\nonumber\\
   &\langle \xi_i (t)\xi_j (0)\rangle = \Big(\delta_{ij}-\frac{p_ip_j}{|{\bf p}|^2}\Big)\,\kappa_T(|{\bf p}|)+\frac{p_ip_j}{|{\bf p}|^2}\,\kappa_L (|{\bf p}|),
\end{align}
where $\eta(p_i)$ is the drag coefficient, $\kappa_{L/T}$ is the longitudinal/transverse momentum diffusion coefficient, and $\xi_i$ denotes the stochastic force acting on the charm quark. We utilized recently estimated $N_f=2+1$ lattice spatial diffusion coefficient $2\pi DT$~\cite{Altenkort:2023oms} to describe the charm quark interaction strength in the medium. The spatial diffusion coefficient can be related to  $\eta(p_i)$ and  $\kappa_{L/T}$~\cite{Moore:2004tg,Kurian:2020orp}.

The current analysis specifically focuses on open charm mesons. We use a combination of fragmentation and heavy-light coalescence mechanisms for the hadronization process. In the fragmentation approach, a charm quark with momentum $p$ fragments into a D-meson with momentum $zp$. We also consider the coalescence model, where the recombination probability is described by the momentum-space Wigner function~\cite{Zhao:2023nrz}.

\begin{figure*}
\centering
\vspace*{1cm}       
\includegraphics[scale=.13]{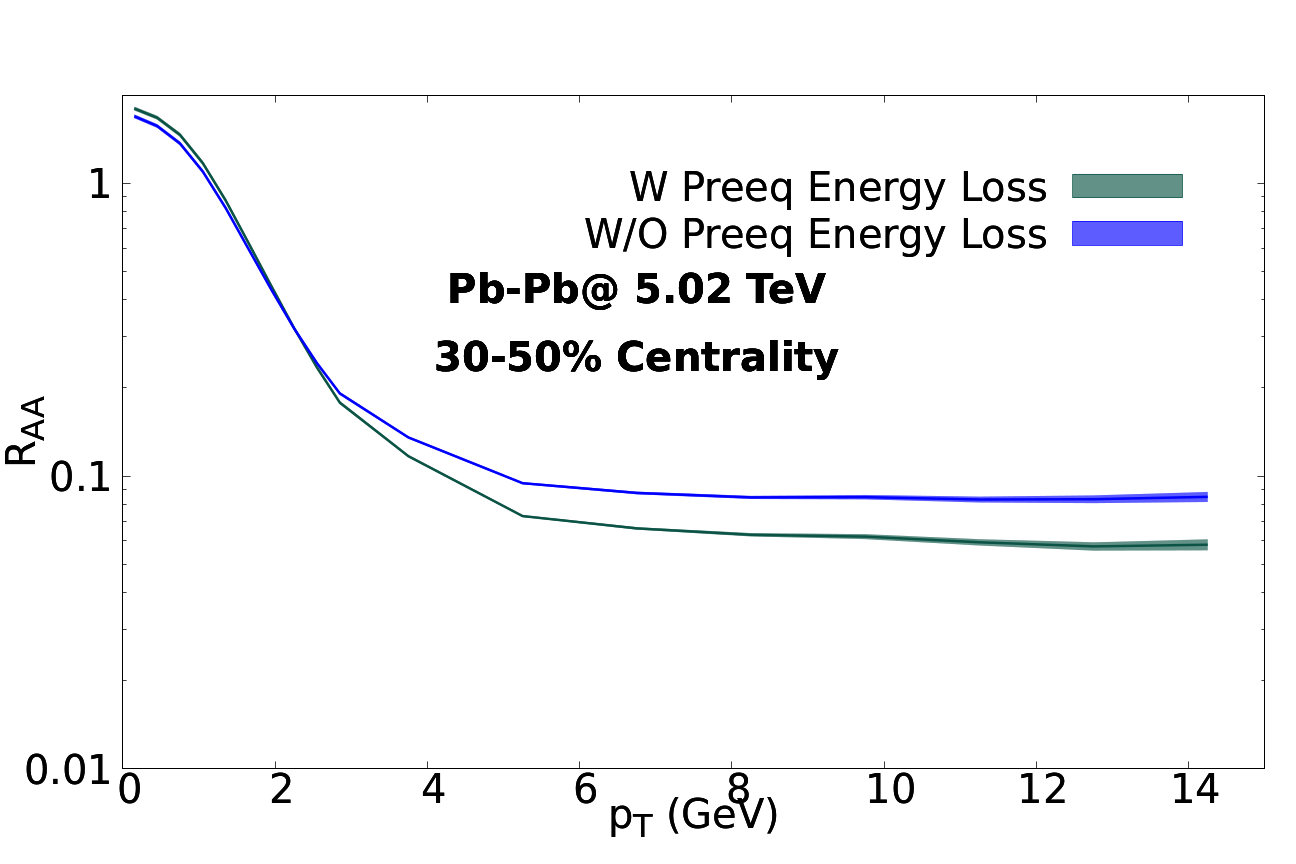}
\includegraphics[scale=.1]{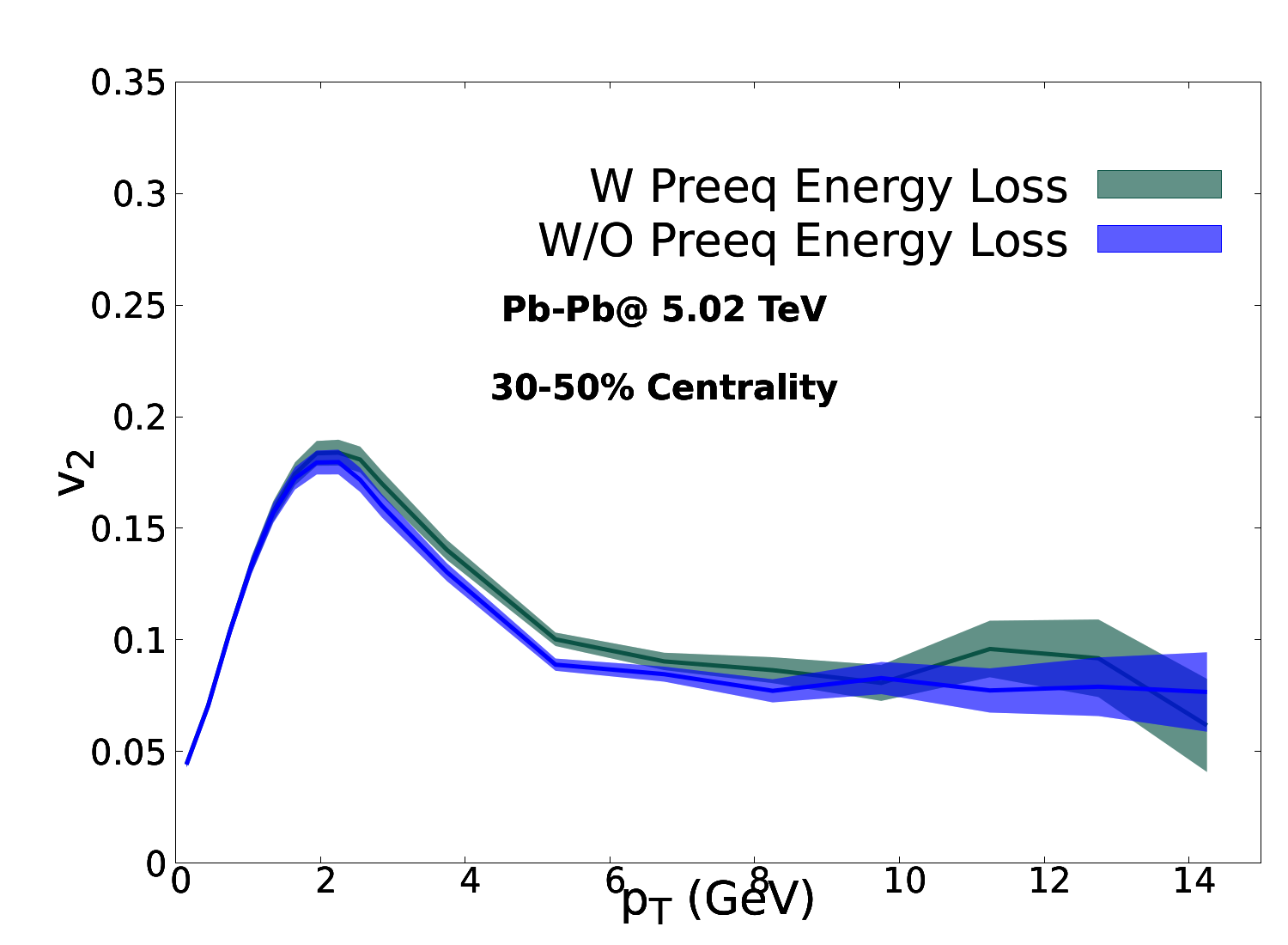}
\caption{The impact of charm quark energy loss in the initial stage of the collision on the D-meson nuclear suppression factor $R_{AA}$ and elliptic flow $v_2$. The temperature-dependent lattice estimation of spatial diffusion coefficient is used to quantify the interaction strength of charm quark in the medium}
\label{fig-2}       
\end{figure*}
To quantify the impact of charm quark evolution in the early stages of the collision, we analyze charm quark observables with and without accounting for its pre-equilibrium energy loss, as shown in figure~\ref{fig-2}. The switching time from the Glasma to plasma phase is fixed at $\tau_w=0.4$ fm. Here, we considered only the fragmentation mechanism for the hadronization process. It is observed that the $R_{AA}$ is modified by the pre-equilibrium energy loss of the charm quark. Elliptic flow is another key observable that measures the azimuthal anisotropy of charm mesons. The elliptic flow, $v_2$, is estimated using the scalar product method. It is shown that the fluctuating initial state modifies the D-meson $v_2$ in two ways. In comparison with a smooth background, fluctuations increase local pressure gradients, which in turn increase $v_2$. Additionally, a fluctuating background causes a decorrelation between the event planes of light and heavy-flavor mesons, which suppresses $v_2$. Refer~\cite{Singh:2023smw} for details. The evolution of charm quarks in the fluctuating initial state further modifies D-meson $v_2$. Despite the relatively short lifetime of the initial phase, we observe that the energy loss of charm quarks in the IP-Glasma phase has a non-negligible effect in the phenomenological study.

\begin{figure}
\centering
\vspace*{1cm}       
\includegraphics[scale=.13]{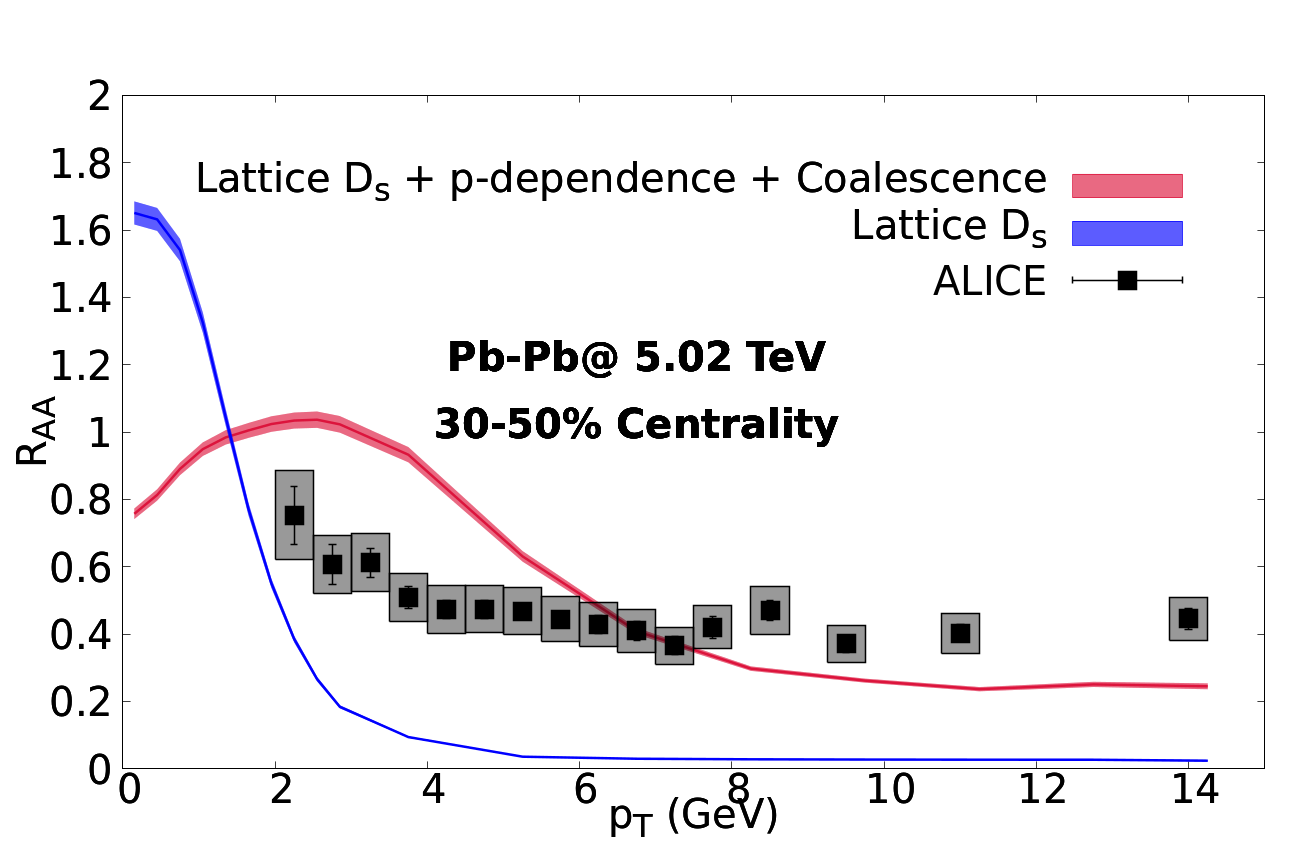}
\caption{Comparison of D meson $R_{AA}$ with the experimental data~\cite{ALICE:2021rxa} for Pb$+$Pb at 5.02 TeV. }
\label{fig-3}       
\end{figure}

In figure~\ref{fig-3}, we consider momentum-dependent heavy quark transport coefficients and the heavy-light coalescence mechanism in addition to the fragmentation process for hadronization. This allows for a comparison of our results with experimental data. A parameterized momentum dependence, motivated by perturbative QCD estimates of heavy quark coefficients, is introduced into the temperature-dependent $2+1$ flavor lattice rates. It is observed that both the momentum dependence and the coalescence hadronization mechanism play significant roles in the phenomenological study of heavy quarks. 

\section{Summary}
One of the interesting questions we aim to address is whether charm quarks can act as efficient probes of the initial phase despite their shorter lifetime and fluctuating energy density profile. An exact solution to this aspect requires a thorough understanding of charm quark interactions with the non-Abelian fields, which is highly non-trivial. In this study, we provide a qualitative estimation of the effect of pre-equilibrium charm quark evolution on charm meson observables by approximating the pre-equilibrium phase as a gluonic system with a temperature. Our analysis shows that charm quarks demonstrate their potential to provide penetrating tomographic information about the initial stage of heavy-ion collisions.\\

\noindent{\bf Acknowledgments:} CG and SJ is supported by the Natural Sciences and Engineering Research Council of Canada (Grant numbers SAPIN-2018-00024 and SAPIN-2020-00048). MK acknowledges the RIKEN SPDR program. BS is supported by the U.S. Department of Energy, Office of Science, Office of Nuclear Physics, under DOE Contract No. DE-SC0012704  and within the framework of the Saturated Glue (SURGE) Topical Theory Collaboration. MS is supported by U.S. DOE Grants No. DE-SC0024347.
%
%
%

\end{document}